\documentclass[prl,twocolumn,nofootinbib,aps,preprintnumbers,amsmath,amssymb,superscriptaddress]{revtex4-1}
\usepackage{graphicx}
\usepackage{color}
\usepackage{bm}
\usepackage{xspace}
\usepackage[normalem]{ulem}
\usepackage{units}
\usepackage{dcolumn}
\usepackage{paralist}
\usepackage{natmove}
\usepackage{natbib}
\usepackage{hyperref}

\begin{document}

\title{Modulation of Semiconductor Superlattice Thermopower Through Symmetry and Strain}
\author{Vitaly~S~Proshchenko}
\affiliation{Ann and H.J. Smead Aerospace Engineering Sciences, University of Colorado Boulder, Boulder, Colorado 80309, USA}
\author{Manoj~Settipalli}
\affiliation{Ann and H.J. Smead Aerospace Engineering Sciences, University of Colorado Boulder, Boulder, Colorado 80309, USA}
\author{Artem~K~Pimachev}
\affiliation{Ann and H.J. Smead Aerospace Engineering Sciences, University of Colorado Boulder, Boulder, Colorado 80309, USA}
\author{Sanghamitra~Neogi}
\email{sanghamitra.neogi@colorado.edu}
\affiliation{Ann and H.J. Smead Aerospace Engineering Sciences, University of Colorado Boulder, Boulder, Colorado 80309, USA}

\date{\today}

\begin{abstract}
In doped semiconductors and metals, the thermopower decreases with increasing carrier concentration, in agreement with the Pisarenko relation. Here, we demonstrate a new strain engineering approach to increase the thermopower of [001] Si/Ge superlattices (SLs) beyond this relation. Using two independent theoretical modeling approaches, we show that new bands form due to the structural symmetry, and, the SL bands are highly tunable with epitaxial substrate strain. The band shifts lead to a modulated thermopower, with a peak $\sim$5-fold enhancement in strained Si/Ge SLs in the high doping regime.
\end{abstract}

\maketitle

Recent advances in nanofabrication and characterization techniques have created exciting opportunities to engineer lattice strain in materials with unprecedented spatiotemporal resolutions. Strain engineering has been demonstrated to enable unique functionalities in materials for a broad range of applications, including optoelectronics~\cite{son2010strain,sanchez2011direct,yahyaoui2014effects,xia2016uniaxial}, electrochemistry~\cite{muralidharan2016strain}, microelectronics~\cite{lee2005strained,lee2015strain,niquet2012effects}, multiferroics~\cite{yang2015strain}, two-dimensional materials~\cite{zhang2015strain,bai2017strain}, functional soft crystals~\cite{wu2016strain}, and more recently, high $T_c$ superconductors~\cite{phan2017effects} and quantum materials~\cite{grosso2017tunable,sohn2018controlling}. Strain is generated due to various mechanisms including lattice mismatch, thermal expansion, phase transition and presence of point or extended defects. In the last two decades, silicon(Si)/germanium(Ge) nanostructures have emerged as key enabling materials in numerous electronic~\cite{thompson200490,meyerson1994high}, optoelectronic~\cite{koester2006germanium,liu2010ge,tsaur1994heterojunction,pearsall1994electronic,engvall1993electroluminescence} and thermoelectric devices~\cite{chen2003recent, dresselhaus2007new, alam2013review,chakraborty2003thermal,lee1997thermal,li2003thermal}, and promising hosts of spin qubits~\cite{shi2011tunable}. The electronic properties of Si/Ge heterostructures witnessed groundbreaking advances with strain engineering~\cite{song2011mobility,boztug2014strained,kasper1995properties,falub2012scaling}. The epitaxial strain due to the nanostructure-substrate lattice mismatch, in particular, has remarkably enhanced the drive current in Si-based devices by controlling the carrier mobilities~\cite{lee2005strained,lee2015strain,niquet2012effects}. 

Epitaxial strain has also been reported to enhance the thermoelectric properties of nanostructured materials~\cite{koga1998carrier,koga1999carrier,koga2000experimental,bahk2010thermoelectric,bahk2012seebeck,geisler2018confinement,geisler2019inducing,nguyen2015enhanced, hinsche2012thermoelectric}. The efficiency of a thermoelectric material is measured by the figure of merit, $ZT = \sigma S^2 T/\kappa$, where $\sigma$ is the electrical conductivity, $T$ is the temperature, $\kappa$ is the thermal conductivity and $S$ is the Seebeck coefficient or thermopower that characterizes the thermoelectric sensitivity of a material. %: $S = \Delta V/\Delta T$, the ratio between the electric voltage ($\Delta V$) established across the material to the temperature difference ($\Delta T$) the material is subjected to. 
Thus, materials with high $S$ are vital for efficient thermoelectric generators and coolers as well as thermal sensors. Several band structure engineering approaches have been proposed to distort the electronic density of states (DOS) and increase $S$ of thermoelectric semiconductors~\cite{heremans2012resonant}. In parallel, nanostructured low-dimensional materials have been shown to increase the energy-dependence of DOS~\cite{cutler1969observation} compared to bulk and improve $ZT$~\cite{mahan1996best,hicks1996experimental, hicks1993effect}. Koga et al. introduced the carrier pocket engineering (CPE) concept to enhance the energy-dependence of the DOS of nanostructured semiconductor superlattices (SLs) %altering the relative carrier contributions from the quantum well and barrier states 
using lattice strain~\cite{koga1998carrier,koga1999carrier,koga2000experimental}. %The states are formed from the minima (pockets) of conduction bands at various high symmetry points of the SL Brillouin zone (BZ). 
Enhanced $S$ at low carrier concentrations is reported in strained III-V semiconductor SLs~\cite{bahk2012seebeck}, %, oxide SLs~\cite{geisler2018confinement,geisler2019inducing} and graphene based materials~\cite{nguyen2015enhanced}.
however, in doped semiconductors and metals, $S$ decreases with increasing the carrier concentration according to the Pisarenko relation~\cite{ioffe1960physics}. Recent first-principles studies tried to optimize $ZT$ of [111] Si/Ge SLs and [001] Si/Ge SL nanowires~\cite{shi2012high} with strain. 
However, the thermopower was shown to follow the Pisarenko relation in strained SLs. Furthermore, the CPE concept was established qualitatively using experimental data and approximated bulk electronic bands~\cite{koga1998carrier,broido1995effect}. The effect of strain on electronic bands of [001] Si/Ge SLs was predicted with first-principles density functional theory (DFT) modeling~\cite{satpathy1988electronic,hybertsen1987theory}, however no connection was made to describe the resulting $S$. A detailed understanding of the relationship between the strain environment in highly technologically relevant [001]-grown Si/Ge heterostructures %fabricated on various substrates, evolution of electronic bands due to the processing related strain and the contribution of these bands to modulate 
and thermopower is still missing.  The strain-electronic transport property relationship will be crucial to fully exploit strain engineering to control electronic properties of future technology-enabling materials.

In this Letter, we present a new strain engineering approach to enhance the thermopower of semiconductor superlattices beyond the Pisarenko relation. We establish the approach by providing a fundamental understanding of the relationship between strain in semiconductor heterostructures and consequential modulation of the thermopower. We investigate the electronic structure and cross-plane transport properties of substrate strained Si$_n$Ge$_m$ SLs, by employing two independent approaches: analytical Kr\"{o}nig-Penny (KP) model~\cite{kittel1996introduction} and first-principles DFT in combination with semi-classical Boltzmann transport equation (BTE). We illustrate the relationship by explaining the following key issues: ({\em i}) the band formation in semiconductor heterostructures due to zone folding and periodic potential, imposed by the structural symmetry; ({\em ii}) the role of in-plane epitaxial substrate strain to control the overall strain environment and tune the valley contributions to modify these newly formed bands; and, lastly, ({\em iii}) the influence of the resulting oscillatory DOS to modulate and enhance thermopower of semiconductor heterostructures in the technologically relevant high-doping regime, deviating from the Pisarenko relation.

\begin{figure*}
\begin{center}
\includegraphics[width=1.0\linewidth]{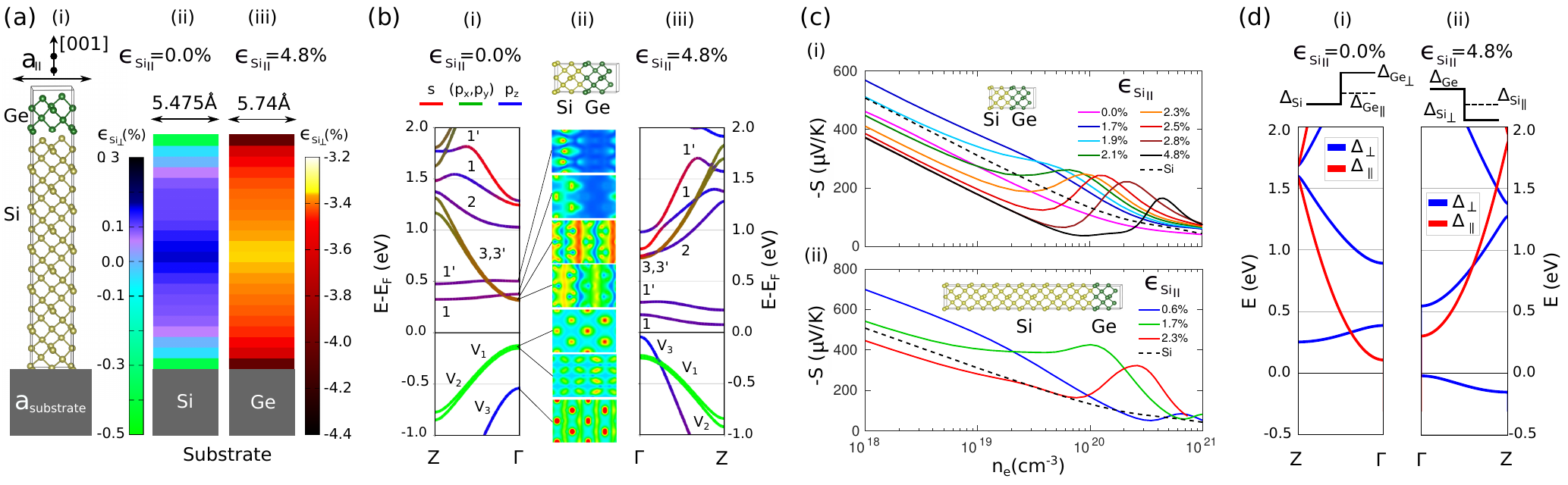}
\caption{{\bf (a) Strain environment in a substrate strained superlattice (SL).} (i) Representative configuration of a Si$_{24}$Ge$_4$ SL studied in this work. ($a_\parallel=a_{\text{substrate}}$) Cross-plane strain ($\epsilon_{\text{Si}\perp}$) in confined Si MLs in a Si$_{24}$Ge$_{4}$ SL grown on (ii) a Si and (iii) a Ge substrate. The in-plane substrate strains ($\epsilon_{\text{Si}\parallel}$) shown at the top of the figures are measured with respect to bulk Si. {\bf (b) Electronic structure properties of a superlattice with substrate strain.} DFT band energies of Si$_4$Ge$_4$ SLs grown on a Si (i) and a Ge (iii) substrate, along the $\Gamma\; - \;  Z$ path of the SL BZ. The red, green, and blue colors represent contributions of the $s$, ($p_x$ and $p_y$), and $p_z$ electrons of Si and Ge to form the bands, respectively. (ii) The charge densities corresponding to the wavefunction of the SL bands close to the Fermi level. {\bf (c) Tunability of electronic transport properties of a superlattice with substrate strain.} Seebeck coefficients of (i) Si$_4$Ge$_4$ SLs and (ii) Si$_{24}$Ge$_4$ SLs grown on substrates with different lattice constants, as a function of carrier concentrations. Inset shows the configurations of the respective SL supercells. The black dashed line represents the bulk Si thermopower. {\bf (d) Kr\"{o}nig-Penney model predictions of the shifts of superlattice electronic bands with substrate strain.} Electronic bands of Si$_4$Ge$_4$ SLs grown on a Si (i) and a Ge (ii) substrate predicted from KP models. The schematic of the strain-split Si and Ge valleys used to construct the KP models, are shown above. The direction of the energy shifts match between the two predictions even though the KP models predict different band energies compared to DFT.}
\label{fig:fig1}
\end{center}
\end{figure*}

Advances in molecular beam epitaxy techniques have enabled pseudomorphic growth of defect-free lattice-mismatched heterostructures~\cite{kuan1991strain,david2018new}, for sufficiently thin layers. The lattice constants parallel to the interface ($a_\parallel$) adjust during growth so that perfectly lattice matched materials can be grown. To accommodate for the mismatch, the lattice constants perpendicular to the interface ($a_\perp$) adjust so that the elastic energy of the two materials is minimized. We simulate the effect of substrate induced strain by modelling the Si$_n$Ge$_m$ supercells with fixed $a_{\parallel} = a_{\text{substrate}}$, and let the supercell relax only in the cross-plane direction ($a_\perp$). We generate the Si$_n$Ge$_m$ tetragonal supercells by replicating a conventional cubic cell (CC) of silicon in the [001] direction and replacing $m$ Si monolayers (MLs) with Ge, since both silicon and germanium have stable FCC diamond lattice structures~\cite{book1, book2}. We model the [001] SLs to periodically extend in the growth direction, to maintain translational invariance. A representative configuration of Si$_n$Ge$_m$ SLs is shown in Fig. \ref{fig:fig1}(a)(i). We investigate two SLs with different numbers of Si ($n$) and Ge ($m$) MLs: (a) $n=m=4$ and (b) $n=24, \;m=4$, grown on different substrates, to discuss the effect of substrate strain on thermopower of SLs for any given period and composition. We have not considered the effects of finite number of periods on thermopower in this analysis. 

To obtain the atomic positions in the strained SLs we performed structural relaxations using DFT. The DFT relaxation method (See Supplemental Materials (SM)) yields the following lattice constants of Si$_4$Ge$_4$ SL: ({\em i}) $a_{\text{Si}\parallel} = a_{\text{Ge}\parallel} = 5.475\;\text{\AA}$,  $a_{\text{Si}\perp}  = 5.45\;\text{\AA}$ and  $a_{\text{Ge}\perp} = 5.94\;\text{\AA}$ on a Si substrate, and, ({\em ii}) $a_{\text{Si}\parallel} = a_{\text{Ge}\parallel} = 5.74\;\text{\AA}$,  $a_{\text{Si}\perp} = 5.26\;\text{\AA}$ and $a_{\text{Ge}\perp} = 5.76\;\text{\AA}$ on a Ge substrate. In contrast with previous studies~\cite{van1986theoretical,hybertsen1987theory,satpathy1988electronic}, DFT relaxation yields $a_{\text{Si}\perp} \neq a_{ \text{Si}\parallel}$ for ({\em i}) the SL on a Si substrate and $a_{\text{Ge}\perp} \neq a_{\text{Ge}\parallel}$ for ({\em ii}) the SL on a Ge substrate. The approach of determining lattice constants by minimizing the macroscopic elastic energy, assumed in previous studies, fails to capture that cross-plane relaxation affects the lattice constants of the entire supercell and not only the mismatched component (({\em i}) Ge or ({\em ii}) Si). In order to characterize the strain environment in the  Si$_n$Ge$_m$ SLs, we estimate the in-plane and the cross-plane strain in MLs by $\epsilon_{i\parallel}= (a_{\parallel}/a_{i}-1)$ and $\epsilon_{i\perp}= (a_{\perp}/a_i-1)$, respectively, with $i=$ Si or Ge~\cite{van1986theoretical} and $a_{\text{Si}}=5.475\;\text{\AA}$ or $a_{\text{Ge}} = 5.74\;\text{\AA}$. Tensile in-plane strain yields $\epsilon_{i\parallel} >0$ while compressive cross-plane strain means $\epsilon_{i\perp} <0$. %In a substrate strained SL, $a_\parallel=a_{\text{substrate}}$, however, the cross-plane strain environment is not uniform. 
Figures~\ref{fig:fig1}(a)(ii)-(iii) display the non-uniform cross-plane ML separations in the confined Si region of a Si$_{24}$Ge$_{4}$ SL, grown on substrates with different $a_\text{substrate}$.  
On an average, the cross-plane MLs are more compressed ($\epsilon_{\text{Si}\perp}\downarrow$) with increasing in-plane tensile substrate strain (Fig.~\ref{fig:fig1}(a)(iii)), as expected. It is evident that the knowledge of both $a_\parallel$ and $a_\perp$(position) is essential to characterize the strain environment in a substrate strained SL. This non-uniform strain environment strongly influences the electronic properties of the SLs. We calculate the electronic structure properties by performing non self-consistent field calculations using the generalized gradient approximation of the exchange-correlation functional by Perdew-Burke-Ernzerhof as implemented in the plane-waves code Quantum Espresso~\cite{giannozzi2009quantum} (See SM) and employ the semi-classical BTE~\cite{ziman1960electrons} as implemented in the BoltzTraP code~\cite{madsen2006boltztrap} to compute thermopower at room temperature~\cite{ashcroftmermin,madsen2006boltztrap}.

We now turn to the main focus of this work which is to demonstrate how substrate strain tunes the thermopower of Si$_n$Ge$_m$ SLs. To this end, we first characterize the effects of structural symmetry and substrate strain on electronic bands. We present the DFT-computed bands of Si$_4$Ge$_4$ SLs on two exemplary substrates, Si (Fig.~\ref{fig:fig1}(b)(i)) and Ge (Fig.~\ref{fig:fig1}(b)(iii)), along the $\Gamma-Z$ path of the tetragonal BZ, corresponding to the cross-plane SL direction. %We refer to the $i^{\text{th}}$ valence state at $\Gamma$ as $V_i$, counting down from the upper valence state. The conduction states are marked with numbers counting up from the lower conduction states. 
The red, green, and blue colors represent the overall contributions from the $s$, $(p_x$, $p_y)$, and $p_z$ electrons (Si and Ge) to the SL bands, respectively. The basic features of the SL bands were described in terms of the zone-folded average bulk bands~\cite{satpathy1988electronic}. Analyzing the electronic bands presented in Fig.~\ref{fig:fig1}(b), we note: (1) The threefold degenerate $p$ states that form the top bulk Si valence band have split into two approximately degenerate $p_x$, $p_y$ states ($V_1, \, V_2$, green) and one nondegenerate $p_z$ state ($V_3$, blue), due to the distortion by the Ge region~\cite{satpathy1988electronic, hybertsen1987theory, tserbak1993unified}. The $p_x,\, p_y$ states ($V_1, \, V_2$) form the valence-band edge in a SL on Si and the $p_z$ state ($V_3$) splits off. The charge density plots in Fig.~\ref{fig:fig1}(b)(ii) show that the $p_x$, $p_y$ states are weakly confined in Ge, while the $p_z$ state is contributed by both the regions. The charge densities in Fig.~\ref{fig:fig1}(b)(ii) correspond to the wave functions of the Si$_4$Ge$_4$ bands shown in Fig.~\ref{fig:fig1}(b)(i) (See SM). The delocalization of the $p_z$ state is caused by the fact that the effective masses of these bands correspond to valence bandwidths that are comparable or larger than the effective barrier potentials between the two regions~\cite{satpathy1988electronic}. The order of these states reverses with increasing the substrate strain, when the SL is on Ge ($a_\parallel\uparrow \rightarrow \epsilon_{i\parallel}\uparrow \rightarrow \epsilon_{i\perp}\downarrow$). The increased $(-)\epsilon_{\perp}$ (Fig.~\ref{fig:fig1}(a)(iii)) causes the $p_z$ band to shift up in energy, while the increased $\epsilon_{\parallel}$, especially in the Ge region, decreases energy levels of the $p_x$, $p_y$ states.

\begin{figure}
\begin{center}
\includegraphics[width=1.0\linewidth]{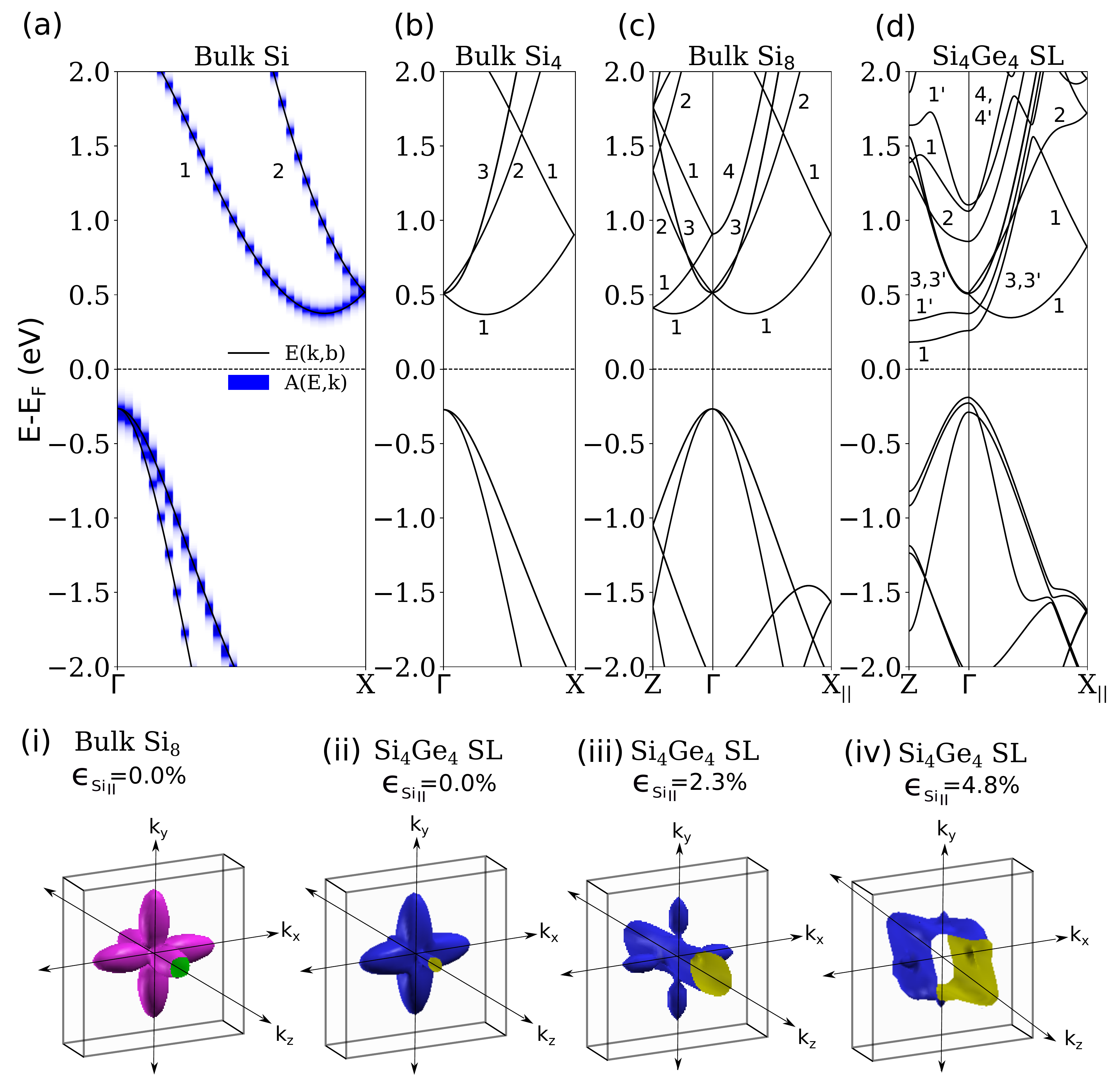}
\caption{\textbf{Electronic bands formation in superlattices and modification with strain}. Bands corresponding to bulk Si lattices with (a) 2 atom, (b) 8 atom, (c) 16 atom in the unit cell, compared to the bands of (d) a strain-symmetrized relaxed Si$_4$Ge$_4$ superlattice. Fermi surfaces shown  correspond to the lowest conduction valleys along $\Gamma-Z$ and $\Gamma-X_\parallel$ directions, represented by type 1 bands of the (i) bulk Si lattice with 16 atom in the unit cell at $\sim$0.5 eV,  (ii)  Si$_4$Ge$_4$ SL grown on Si substrate at $\sim$0.5 eV, (iii) strain-symmetrized relaxed Si$_4$Ge$_4$ SL at $\sim$0.5 eV, and (iv)  Si$_4$Ge$_4$ SL grown on Ge substrate at $\sim$0.6 eV.}
\label{fig:fig2}
\end{center}
\end{figure}

(2) When layered with Ge, the six Si conduction $\Delta$ valleys, located near the $X$-points of the diamond BZ, become inequivalent. As we illustrate in detail later, these $k$-space valleys are zone folded close to the $\Gamma$ point due to the supercell BZ periodicity (see Fig.~\ref{fig:fig2}). The folding results in two [$s,\, p_z$] states ($1, \, 1^\prime$, red-blue) and two two-fold degenerate [$s,\, p_x,\, p_y$] states ($3, \, 3^\prime$, red-green). The transverse valleys parallel to the layers introduce a small effective barrier potential between the Si and the Ge region~\cite{satpathy1988electronic} causing the ``in-plane" [$s,\, p_x,\, p_y$] states to be delocalized. The ``cross-plane" [$s,\, p_z$] states are strongly confined in Si due to the large barrier of the longitudinal valleys, and, are minimally dispersive~\cite{tserbak1993unified}. The minigap-splitting between the [$s,\, p_z$] states is due to the potential barrier and the intervalley mixing effects. When the SL is on Si, the conduction band edge is formed by the ($3, \, 3^\prime$) states. The ($1, \, 1^\prime$) states are higher in energy due to confinement and strain effects. The order of the [$s,\, p_x,\, p_y$] and [$s,\, p_z]$ states reverse when the SL is on Ge, in an analogous manner to the valence states. In general, the tensile substrate strain in the Si zone causes the [$s,\, p_x,\, p_y$] states to be higher in energy than the [$s,\, p_z$] states. Figure~\ref{fig:fig1}(b) illustrates that substrate strain dictates the relative energy of the [$s,\, p_x,\, p_y$] and [$s,\, p_z$] states and creates a gap in the SL conduction zone. 

(3) The resulting non-monotonic electronic DOS leads to oscillations in the electronic transport coefficients. Fig.~\ref{fig:fig1}(c) shows the thermopower of two substrate strained $n$-type SLs. Similar oscillatory thermopower has been reported for III-V SLs~\cite{vashaee2006cross, bian2007cross, bahk2012seebeck} but not for Si/Ge heterostructures, the widely used systems for numerous applications, and no relationship has been established between the strain-induced band shifts and the oscillatory $S$. Furthermore, the clear deviations of $S$ of strained Si/Ge SLs from the Pisarenko relation have never been reported. Our results establish a new strain engineering approach to increase $S$ beyond this relation, orthogonal to the proposed approaches to distort the DOS by adding impurities to the thermoelectric materials ~\cite{heremans2012resonant}. The oscillations in $S$ directly correspond to the gap between the [$s,\, p_x,\, p_y$] and the [$s,\, p_z$] states in the conduction zone. Therefore the thermopower at different carrier concentrations can be optimized by tuning this gap. Figure~\ref{fig:fig1}(c)(i) illustrates this remarkable tunability of $S$ of Si$_4$Ge$_4$ SLs grown on substrates inducing strains ranging from $\epsilon_{\text{Si}\parallel}=0.0\%$ (Si substrate, magenta) to $4.8\%$ (Ge substrate, black solid). Figure~\ref{fig:fig1}(c)(ii) displays similar oscillations in $S$ of substrate strained Si$_{24}$Ge$_{4}$ SLs with $\epsilon_{\text{Si}\parallel}=0.6\%$ (blue), $1.7\%$ (green), and $2.3\%$ (red), respectively. The black dashed line represents bulk Si $S$ at different carrier concentrations. %In strain-symmetrized Si$_4$Ge$_4$ and Si$_{24}$Ge$_{4}$ SLs, with $2.3\%$ (orange) and $0.6\%$ strains (blue), respectively, the Si regions are under biaxial tensile strain. 
A general trend can be noted from Fig.~\ref{fig:fig1}(c) that a smaller tensile substrate strain improves $S$ at smaller carrier concentrations, while higher strain improves $S$ at higher carrier concentrations. For example, $0.6\%$ substrate strain improves the $S$ of a Si$_{24}$Ge$_{4}$ SL at $n_e\leqslant 2\times 10^{19} \;$cm$^{-3}$, while higher $1.7\%$ and $2.3\%$ strains result in a 2.5 times higher $S = 426\;\mu$V/K at $n_e=1\times 10^{20} \;$cm$^{-3}$ and a $5.2$ times higher $S = 323\;\mu$V/K at $n_e=2.6\times 10^{20} \;$cm$^{-3}$, than the corresponding values of a strain-symmetrized SL ($0.6\%$ strain (blue)), respectively. The peak values in these two cases are $3.2$ and $3.9$ times higher than the corresponding $S$ values of bulk Si at the same carrier concentrations, respectively.

In the previous paragraphs we illustrated how substrate strain tunes Si$_n$Ge$_m$ SL bands and control thermopower, through the DFT results. We now provide an independent demonstration of the physical phenomena, using the KP model~\cite{zachai1990photoluminescence,koga1999carrier,koga1998carrier,bahk2012seebeck}. We construct the KP models using the effective well and barrier potentials between the strain-split Si and Ge $\Delta$ valleys (Fig.~\ref{fig:fig1}(d)) and their effective masses~\cite{zachai1990photoluminescence,koga1999carrier} (See SM). The bands shown in Fig.~\ref{fig:fig1}(d) are with respect to the unstrained Si $\Delta$ valley energy level. Intervalley mixing effects are ignored in these calculations. Figure~\ref{fig:fig1}(d)(i)-(ii) illustrates that tensile substrate strain shifts the conduction bands corresponding to $\Delta_\parallel$ and $\Delta_\perp$ valleys in opposite directions to create a gap, in a manner very similar to the DFT prediction shown in Fig. \ref{fig:fig1}(b). This further corroborates the concept that Si$_n$Ge$_m$ SL minibands can be strain engineered to modulate the thermopower. 

To further elucidate the fundamental relationship between the band shifts and the tunable $S$ (Fig.~\ref{fig:fig1}), we analyze the Si$_4$Ge$_4$ SL bands and compare with the zone-folded bulk Si bands. In Fig.~\ref{fig:fig2}(a) we show bulk Si bands along $\Delta$ ($\Gamma (0,0,0)\rightarrow X (0,0,2\pi/a)$) of the diamond BZ of a two atom primitive unit cell. In a cubic BZ of an eight atom conventional unit cell (Si$_4$), the $X$ points $[(\pm 2\pi/a,0,0),$ $(0,\pm 2\pi/a,0)$ and $(0,0,\pm 2\pi/a)]$ are folded to the $\Gamma$ point. A new set of bands 3 appear at the energy level where bands 1 intersect the $\Gamma$ point. We discuss the origin of bands 3 by analyzing the bands of a [001] replicated sixteen-atom supercell (Si$_8$). This supercell BZ matches with the SL periodicity and facilitates a direct comparison between the SL and the folded bulk bands. The $(0,0,\pm \pi/a)$ points of the cubic BZ are folded to the $\Gamma$ point of the tetragonal BZ, while the $X_\parallel$ $[(\pm \pi/a,0,0)$ and $(0,\pm \pi/a,0)]$ points remain unaltered. The overlap of folded $\Gamma-Z$ bands at the $\Gamma$ point introduces a new set of bands 4 along $\Gamma-X_\parallel$ (Fig.~\ref{fig:fig2}(c)). The Fermi surfaces (FS) with valleys centered at $\sim(0.15)(2\pi/a)$ from the $\Gamma$ point evolve uniformly in all directions and intersect the $\Gamma$ point at $\sim$0.5 eV (Fig.~\ref{fig:fig2}(i)). The overlap of the four $\Delta_\parallel$ valleys at $\sim$0.5 eV and the two $\Delta_\perp$ valleys at the same energy level results in the bands 3 along $\Gamma-Z$ and $\Gamma-X_\parallel$, respectively. Similarly, bands 4 along $\Gamma-X_\parallel$ originate at the energy level where folded $\Gamma-Z$  bands from $\Delta_\perp$ valleys intersect the $\Gamma$ point. 

The SL bands mostly retain Si like nature, however, display band splittings and gaps opening at the BZ boundaries (Fig.~\ref{fig:fig2}(d)) due to the potential perturbation and the intervalley mixing effects. The bands 1 along $\Gamma-Z$ yield a pair of slightly slit, minimally dispersive bands, as discussed before. The FS corresponding to the bands 1 are considerably more flat along the $\Gamma-Z$ direction than in $\Gamma-X_\parallel$ direction (Fig.~\ref{fig:fig2}(iii)). As a result, the bands (3, $3^\prime$) originate at different energy levels between the two directions. The appearance of new bands in the cross-plane direction in Si$_n$Ge$_m$ SLs due to the overlap of in-plane bands folded at the $\Gamma$ point has not been reported~\cite{satpathy1988electronic,van1986theoretical,hybertsen1987theory,tserbak1993unified}. With increase of in-plane strain $\epsilon_{Si\parallel}$, the conduction band minima shifts from in-plane $\Gamma-X_\parallel$ to cross-plane $\Gamma-Z$ miniband valleys. A corresponding change in the relative evolution of FS with respect to the Fermi energy (Fig.~\ref{fig:fig2}(ii)-(iv)) can be noted. The relative movement of the bands modulates the thermopower of Si$_4$Ge$_4$ SLs, as illustrated in Fig.~\ref{fig:fig1}(c). 

Here we demonstrate the direct relationship between the modified bands and the resulting cross-plane thermopower of substrate strained Si/Ge SLs. New bands appear due to the interaction between the zone-folded bulk Si bands, the folding dictated by SL structural symmetry. We illustrate that the SL bands are remarkably tunable with epitaxial substrate strain, using two independent modeling approaches. The resulting non-monotonic DOS modulates the thermopower at different carrier concentrations. %by tuning the conduction bands with controlled fabrication processes. 
We report enhanced thermopower of substrate strained SLs at high carrier concentrations, over the Pisarenko relation. For example, a $2.3\%$ substrate strained Si$_{24}$Ge$_{4}$ SL has a peak $S = 323\;\mu$V/K at $n_e=2.6\times10^{20} \;$cm$^{-3}$, 5.2 and 4 times higher than a strain-symmetrized SL and the bulk Si, respectively. We anticipate that our study will encourage future investigations to enhance thermoelectric properties of a broad class of strain-engineered semiconductor superlattices in the high-doping regime. Additionally, the insight will help to develop a new approach to estimate thermally induced strain in electronic devices during operation by monitoring thermopower, and to prevent failure due to thermally induced mechanical stresses. It is expected that the fundamental understanding will help to exploit strain engineering strategies on a class of future technology-enabling layered materials, including van der Waals heterostructures.

\section{Acknowledgements}
We thank Davide Donadio for critical reading of the manuscript. The work is funded by the Defense Advanced Research Projects Agency (Defense Sciences Office) [Agreement No.: HR0011-16-2-0043]. All computations were performed using the Extreme Science and Engineering Discovery Environment (XSEDE), which is supported by National Science Foundation grant number ACI-1548562. 
V. S. P. and M. S. contributed equally to this work.

\bibliography{strainLiterature}

\section{Supplementary Materials}
{\bf DFT method details:} The supercell relaxation and subsequent electronic structure calculations is performed with DFT using the generalized gradient approximation (GGA) of the exchange-correlation functional by Pedrew-Burke-Ernzerhof (PBE)~\cite{perdew1996generalized} as implemented in the plane-waves code Quantum Espresso (QE)~\cite{giannozzi2009quantum}. We employ scalar relativistic normconserving pseudopotentials for both Si and Ge atoms~\cite{giannozzi2009quantum}. The Kohn-Sham orbitals expanded in terms of a plane wave basis set, had a cutoff energy of 30 Ry for all calculations, to accurately calculate electronic states~\cite{van1986theoretical,satpathy1988electronic}. A convergence threshold for self-consistency was chosen to be $10^{-9}$. We haven't included spin-orbit interaction in our analysis since the magnitude of the strain energy level splittings were shown to be larger than the spin-orbit splittings~\cite{hybertsen1987theory}. The cross-plane lattice constants and the atomic positions in Si$_n$Ge$_m$ SLs are optimized using Broyden-Fletcher-Goldfarb-Shanno Quasi-Newton algorithm, sampling the BZ with $4\times4\times 4$ and $4\times 4\times 2$ $k$-point mesh for Si$_4$Ge$_4$ and Si$_{24}$Ge$_4$ SL, respectively. The $k$-mesh is generated using Monkhorst-Pack scheme~\cite{monkhorst1976special}. As a reference, the bulk Si and Ge lattice constants obtained with similar method are $5.475\;\text{\AA}$ and $5.74\;\text{\AA}$, respectively, $\sim 1\%$ higher than experimental values~\cite{semiconductor2012general}.

We perform non self-consistent field (NSCF) calculations to obtain the electronic band energies of Si$_4$Ge$_4$ and Si$_{24}$Ge$_4$ SL, using a dense $k$-point mesh with $\sim 50,000$ and $\sim 20,000$ points in the BZ, respectively. Such sampling is necessary to converge the calculation of the electronic transport coefficients.

The band structures and charge densities presented in Fig.1(b) in the main manuscript are obtained using the vienna ab initio simulation package (VASP)~\cite{kresse1996efficiency,kresse1996efficient} with the PBE exchange-correlation functional\cite{perdew1996generalized} and projector augmented plane-wave (PAW) pseudopotentials~\cite{kresse1999ultrasoft,blochl1994projector} and visualized with VESTA~\cite{momma2011vesta} program. The $s$, $p$, and $d$ wavefunction characters of each band were calculated by projecting the wavefunctions onto spherical harmonics within spheres of a radii $2.48$ \text{\AA} and $2.30$ \text{\AA} for Si and Ge, respectively.

{\bf Kr\"{o}nig-Penney (KP) model construction:} Two sets of KP models were constructed, one corresponding to the $\Delta_\parallel$ valleys and the other corresponding to the $\Delta_\perp$ valleys. We used previously reported deformation potentials to compute the valley splittings due to strain~\cite{koga1999carrier}. 
The effective barrier potentials of KP models of a Si$_4$Ge$_4$ SL, corresponding to $\Delta_\parallel$ \& $\Delta_\perp$ valleys, are found to be 0.14 eV \& 0.78 eV (on Si substrate) and 0.09 eV \& 0.89 eV (on Ge substrate), respectively~\cite{koga1999carrier}. The Si valleys form the well regions while the Ge valleys form the barrier regions in both the strain cases. The effective masses for the $\Delta_\parallel$ \& $\Delta_\perp$ KP models are taken to be $0.19m_e$ and $0.92m_e$, for both strain cases,
respectively~\cite{rieger1993electronic}.

\end{document}